\documentclass[sigconf]{acmart}

\usepackage{amsmath}

\usepackage{xspace}
\usepackage{tabularx}
\usepackage{makecell}
\usepackage{subcaption}
\usepackage{enumitem}
\usepackage{pifont}
\usepackage{multirow}
\usepackage{booktabs}
\usepackage{graphicx}
\usepackage{etoolbox}
\usepackage[all]{nowidow}
\usepackage{colortbl}
\usepackage{latexsym}
\usepackage{svg}
\usepackage{rotating}
\usepackage{algorithm}
\usepackage[noend]{algpseudocode}

\usepackage{balance}

\newenvironment{myproof}{{\noindent\it Proof.}}{\hfill $\square$\par}

\usepackage{tikz}

\newcommand{\blackcircle}[1]{%
  \begin{tikzpicture}[baseline=(char.base)]
    \node[shape=circle,fill=black,text=white,inner sep=0.5pt] (char) {#1};
  \end{tikzpicture}%
}

\newcommand{\parabf}[1]{\medskip\noindent\textbf{#1}}

\newcommand{\paraf}[1]{\noindent\textbf{#1}}

\newcommand{\revise}[1]{{#1}}

\newcommand{\sysname}{\texttt{WebANNS}\xspace}

\AtBeginDocument{%
  \providecommand\BibTeX{{%
    \normalfont B\kern-0.5em{\scshape i\kern-0.25em b}\kern-0.8em\TeX}}}

\copyrightyear{2025}
\acmYear{2025}
\setcopyright{acmlicensed}
\acmConference[SIGIR '25]{Proceedings of the 48th International ACM SIGIR Conference on Research and Development in Information Retrieval}{July 13--18, 2025}{Padua, Italy}
\acmBooktitle{Proceedings of the 48th International ACM SIGIR Conference on Research and Development in Information Retrieval (SIGIR '25), July 13--18, 2025, Padua, Italy}
\acmDOI{10.1145/3726302.3730115}
\acmISBN{979-8-4007-1592-1/2025/07}

\begin{document}

\title{\sysname: Fast and Efficient Approximate Nearest Neighbor Search in Web Browsers}

\author{Mugeng Liu}
\affiliation{%
  \institution{School of Computer Science,\\ Peking University}
  \city{Beijing}
  \country{China}}
\email{lmg@pku.edu.cn}

\author{Siqi Zhong}
\affiliation{%
  \institution{Fudan University}
  \city{Shanghai}
  \country{China}}
\email{sqzhong21@m.fudan.edu.cn}

\author{Qi Yang}
\affiliation{%
  \institution{Institute for Artificial Intelligence, Peking University}
  \city{Beijing}
  \country{China}}
\email{qi.yang@stu.pku.edu.cn}

\author{Yudong Han}
\affiliation{%
  \institution{Institute for Artificial Intelligence, Peking University}
  \city{Beijing}
  \country{China}}
\email{hanyd@pku.edu.cn}

\author{Xuanzhe Liu}
\affiliation{%
  \institution{School of Computer Science,\\ Peking University}
  \city{Beijing}
  \country{China}}
\email{xzl@pku.edu.cn}

\author{Yun Ma}
\authornote{Corresponding author.}
\affiliation{%
  \institution{Institute for Artificial Intelligence, Peking University}
  \city{Beijing}
  \country{China}}
\email{mayun@pku.edu.cn}

\renewcommand{\shortauthors}{Mugeng Liu et al.}

\begin{abstract}
Approximate nearest neighbor search (ANNS) has become vital to modern AI infrastructure, particularly in retrieval-augmented generation (RAG) applications. 
Numerous in-browser ANNS engines have emerged to seamlessly integrate with popular LLM-based web applications, while addressing privacy protection and challenges of heterogeneous device deployments.
However, web browsers present unique challenges for ANNS, including computational limitations, external storage access issues, and memory utilization constraints, which state-of-the-art (SOTA) solutions fail to address comprehensively.

We propose \sysname, a novel ANNS engine specifically designed for web browsers. 
\sysname leverages WebAssembly to overcome computational bottlenecks, designs a lazy loading strategy to optimize data retrieval from external storage, and applies a heuristic approach to reduce memory usage.
Experiments show that \sysname is fast and memory efficient, achieving up to $743.8\times$ improvement in 99th percentile query latency over the SOTA engine, while reducing memory usage by up to 39\%.
Note that \sysname decreases query time from 10 seconds to the 10-millisecond range in browsers, making in-browser ANNS practical with user-acceptable latency.
\end{abstract}



\maketitle


\vspace{-2em}
\section{Introduction}

Approximate Nearest Neighbor Search (ANNS) has become a vital component of modern artificial intelligence infrastructure~\cite{tian2024fusionanns}, with wide-ranging applications such as retrieval-augmented generation (RAG)~\cite{semnani2023wikichat, zhoudocprompting, siriwardhana2023improving}, data mining~\cite{tagami2017annexml}, search engines~\cite{zhang2022uni}, and recommendation systems~\cite{zhang2018visual,huang2020embedding}. 
ANNS aims to find the top-\textit{k} most similar vectors in large-scale, high-dimensional search spaces given a query vector. 
In the growing application landscape of ANNS, the query speed directly affects response latency, potentially becoming a performance bottleneck and impacting user experience~\cite{jin2024ragcache}.

Among ANNS engine implementations, in-browser ANNS engines~\cite{wang2024mememo,voy,semanticfinder} are emerging to integrate with popular LLM-based web applications (web apps)~\cite{mememo_app}.
These web apps combine in-browser ANNS with cloud-based web services to enhance service quality by leveraging personalized user data.
In-browser ANNS engines offer privacy guarantees by retrieving user data locally and only sending essential data to cloud-based web services, particularly useful in sensitive domains like finance, education, and healthcare~\cite{chung2023challenges,wutschitz2023rethinking, ghodratnama2023adapting}.
Furthermore, leveraging the cross-device, out-of-the-box nature of the web, these engines overcome the challenge of integrating vector databases across diverse heterogeneous user devices, which otherwise incurs significant development costs and raises deployment and usage barriers~\cite{draxler2023gender,zamfirescu2023johnny}.

Despite the advantages, web browsers propose unique and significant limitations to the compute-intensive and memory-hungry nature of ANNS.
The SOTA in-browser ANNS engine is Mememo~\cite{wang2024mememo}, published in \textit{SIGIR'24}.
Mememo uses the hierarchical navigable small world (HNSW) algorithm, which is SOTA regarding construction and query efficiency~\cite{hnsw,wang2024mememo}.
HNSW is widely used among real-world retrieval and RAG toolkits including FAISS~\cite{douze2024faiss}, Pyserini~\cite{lin2021pyserini}, PGVector~\cite{pgvector}, and LangChain~\cite{langchain}.
Moreover, Mememo extends browser memory via IndexedDB API~\cite{indexeddb} to store data on device disks and reduces query overhead through data pre-fetching.

However, Mememo fails to thoroughly explore browser limitations on ANNS, thereby overlooking critical optimization opportunities and potentially leading to unacceptable query latency that extends into the sub-minute range.
Therefore, we conduct a comprehensive measurement of Mememo, identifying three key limitations of SOTA in-browser ANNS engines.

\paraf{Computational performance.}
Browsers prioritize cross-device compatibility and development flexibility over computational performance, heavily relying on interpreted languages like JavaScript for computation. 
This reliance leads to significant overhead in similarity calculations and sorting operations among vectors during ANNS processes. 
Measured results show that the overhead for a single query, excluding external storage access time, exceeds 100ms, with similarity calculations accounting for over 40\%, which highlights the high computational cost in browsers.

\paraf{External storage access.}
Consecutive accesses to IndexedDB are extremely slow~\cite{indexeddb_slow}.
To minimize IndexedDB access frequency, Mememo heuristically pre-fetches additional vectors that might be queried in a single access.
Our measurement indicates that the prefetch mechanism may fetch over 80\% irrelevant vectors, causing performance degradation when heuristics fail.

\paraf{Restricted memory utilization.}
In-browser ANNS shares memory resources with other functionalities like user interface interactions. 
Excessive memory usage during queries can make the browser unresponsive.
Mememo uses predefined fixed cache sizes for queries, which cannot adapt to the varying resource conditions of devices and browsers, resulting in inefficient use of limited memory resources.

To address these limitations, we propose \sysname, a fast and memory-efficient ANNS engine specifically designed for browsers, which incorporates three key designs.

First, \sysname utilizes WebAssembly (Wasm)~\cite{wasm} to accelerate the computation and sorting operations in browsers. 
Wasm is a binary instruction format that enables near-native computing speed in browsers, offering enhanced performance over JavaScript for compute-intensive tasks.
However, Wasm is limited by only 32-bit addressing~\cite{wasm_memory_limit} and sandbox isolation, resulting in constrained memory capacity and no direct access to external storage.
To address this challenge, \sysname utilizes a three-tier data management mechanism involving Wasm, JavaScript, and IndexedDB. 
JavaScript acts as an intermediary between Wasm and IndexedDB, serving as a cache layer, a data exchange hub, and a bridge linking Wasm's synchronous execution model and IndexedDB's asynchronous model.

Second, \sysname employs a lazy external storage access strategy to reduce IndexedDB access frequency and avoid unnecessary data loading.
However, the completely lazy loading strategy may break the data dependency in the query process of HNSW, where the similarity of a queried vector must be calculated to decide whether neighboring vectors should be queried.
This disruption may lead to incorrect query paths and redundant calculations.
To address this challenge, \sysname employs a phased lazy loading strategy to efficiently load data while ensuring the correct query path.

Third, \sysname uses a heuristic approach to estimating the minimum memory required without impacting query latency, reducing unnecessary memory usage and minimizing effects on other browser functionalities.
However, finding the optimal memory usage across different datasets, browsers, and devices is challenging.
To address this challenge, \sysname models the relationship between query latency and memory size, reducing memory iteratively until latency exceeds a set threshold.

We develop a prototype of \sysname and evaluate it using five commonly used datasets, ranging in size from 4MB to 7.5GB.
Experiments are conducted across three mainstream browsers and three different devices.
Experiment results demonstrate that \sysname achieves up to a $743.8\times$ improvement in 99th percentile (P99) query latency compared to Mememo, while reducing memory usage by up to 39\%.
Moreover, \sysname can reduce query latency from 10 seconds to just 10 milliseconds in a browser, making in-browser ANNS feasible with user-acceptable latency.

In summary, this paper makes the following key contributions\footnote{\url{https://github.com/morgen52/webanns}}

\begin{itemize}
    \item We conduct a measurement study on the SOTA in-browser ANNS engine and identify three major limitations imposed by browsers on ANNS.
    \item We propose \sysname, a novel ANNS engine specifically for browsers, which leverages Wasm for computational efficiency, designs a lazy loading strategy for efficient disk access, and applies a heuristic approach to reduce memory footprint.
    \item We develop a prototype of \sysname and conduct comprehensive evaluations across diverse datasets, browsers, and devices, indicating the significant enhancement of \sysname.
\end{itemize}

\section{Background and Motivation}

This section introduces the background of in-browser ANNS and our measurement study on the SOTA engine.

\subsection{In-browser ANNS}

Integrating personalized data into web apps to enhance service quality is increasingly popular. 
For example, the ChatGPT web app~\cite{chatgpt} can remember user information with permission to generate personalized responses. 
To protect privacy, numerous web apps~\cite{mememo_app} are integrating in-browser ANNS with cloud-based web services, keeping sensitive data within the user's browser instead of on potentially untrustworthy cloud servers. 
Additionally, utilizing the cross-device, out-of-the-box nature of the web, in-browser ANNS engines enable easy multi-device access with one-time development effort.

\subsubsection{Indexing Algorithm}

Given a query vector, ANNS algorithms use similarity metrics, such as Euclidean distance, to identify the top-\textit{k} nearest neighbors from high-dimensional, large-scale datasets. 
To reduce computational costs, most ANNS algorithms~\cite{artem2015invertedmultiindex, chen2021spann, zhang2023fast, zili2024fastvector} employ indexing algorithms to prune data regions unlikely to contain nearest neighbors. 

Graph-based indexing is popular in in-browser ANNS~\cite{wang2024mememo,voy,semanticfinder} because it delivers high accuracy and low latency with limited computational resources.
For example, the hierarchical navigable small world (HNSW) algorithm serves as the indexing backbone of the SOTA in-browser ANNS engine, Mememo, published in \emph{SIGIR'24}.
Known for its SOTA construction and querying performance among graph-based indexing algorithms~\cite{hnsw}, HNSW is widely used in real-world retrieval and RAG toolkits like FAISS~\cite{douze2024faiss}, Pyserini~\cite{lin2021pyserini}, PGVector~\cite{pgvector}, and LangChain~\cite{langchain}.

HNSW constructs a multi-layer graph, with each layer forming a navigable small-world network.
The key idea is to use these hierarchical layers to efficiently navigate high-dimensional data, narrowing down the search space as the query descends through the layers. 
The query process starts at the top layer with a greedy algorithm that iteratively selects the nearest neighbor until no closer vectors are found.
This process quickly eliminates distant vectors, significantly reducing computational load.
As the search descends to lower layers, the algorithm continues to refine its list of nearest neighbors. 
Each subsequent layer is more densely connected than the previous one, which allows for a more precise local search. 
This hierarchical graph effectively balances the trade-off between search accuracy and computational cost.

\subsubsection{Optimization of SOTA In-browser ANNS Engine}

To mitigate the constraints of browsers, SOTA in-browser ANNS engine, Mememo, introduces two browser-specific optimizations.

\paraf{Storage expansion}. Web pages can encounter memory limits as low as 256MB~\cite{memory_limit}, allowing storage of only up to 83,000 vectors of 384 dimensions, excluding other memory usage. 
Moreover, browsers restrict access to the OS file system for security, preventing direct data storage on disks. 
Mememo uses IndexedDB, a cross-browser key-value storage utilizing up to 80\% of client disk space~\cite{indexeddb_description}. 
IndexedDB is a low-level API for client-side storage of significant amounts of structured data. It is the major standard interface in browsers for large dataset storage~\cite{indexeddb}.

\paraf{Prefetching for efficient disk access}. 
While IndexedDB mitigates memory constraints, browser sandboxing can cause extremely slow data access during consecutive IndexedDB API calls~\cite{indexeddb_slow}. 
To reduce disk access frequency, when vector values are missing from the memory during a query, Mememo prefetches the current layer's $p$ neighbors from IndexedDB into memory, where $p$ is the pre-defined cache size in memory.

\subsection{Limitations of In-browser ANNS}

We conduct a measurement study on the SOTA in-browser ANNS engine, Mememo, and identify three key limitations.

\parabf{Setup.} We utilize a widely adopted dataset~\cite{jin2024ragcache} from the Wikipedia corpus~\cite{wiki_dataset}, consisting of approximately $ 0.3$ million documents from the most popular Wikipedia pages. 
The 7.5GB dataset, named Wiki-480k, comprises 480,000 text items with 768-dimensional embeddings.
Experiments are conducted in Chrome on a Linux system equipped with an Intel Core i5-13400F processor (13th generation), performing 100 queries with random vectors. 
We measure P99 query latency for worst-case and average latency for typical performance.

\subsubsection{High Computational Overhead}

\begin{figure}[t!]
\centering
\begin{subfigure}[t]{0.22\textwidth}
    \centering
    \includegraphics[width=\textwidth]{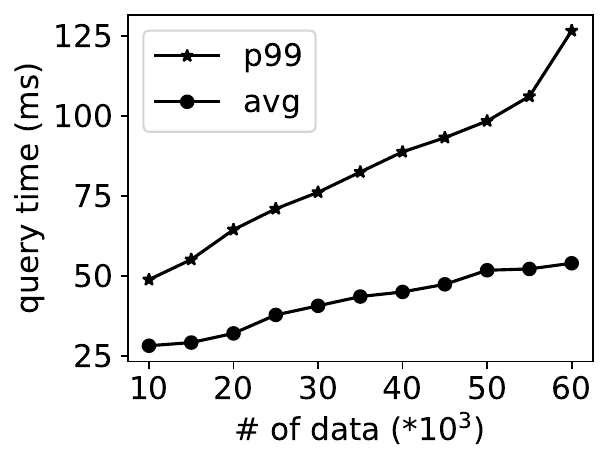}
    \caption{Computational latency excluding IndexedDB access.}
    \label{fig:js_computing1}
\end{subfigure}
\hfill
\begin{subfigure}[t]{0.22\textwidth}
    \centering
    \includegraphics[width=\textwidth]{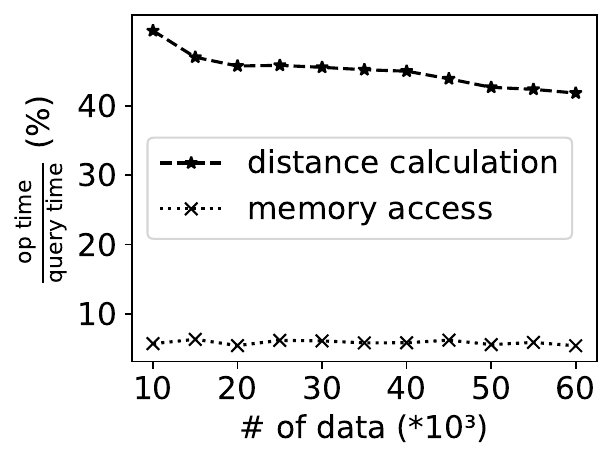}
    \caption{Computational bottlenecks.}
    \label{fig:js_computing2}
\end{subfigure}
\vspace{-1em}
\caption{Computing overhead and bottlenecks.}
\vspace{-1.5em}
\end{figure}

We gradually increase the size of the uploaded data from Wiki-480k and measure the query latency, excluding IndexedDB access latency, under conditions of unlimited memory. 
When the data size exceeds 60,000 items (over 901MB), which accounts for only 12\% of the size of Wiki-480k, the webpage of Mememo crashes due to out-of-memory errors during query process.

Fig.~\ref{fig:js_computing1} illustrates that computational latency generally exceeds 50ms and can surpass 125ms.
The latency becomes a bottleneck for in-browser ANNS, especially in RAG web apps where LLM web service can generate tokens in under 100ms~\cite{tian2024fusionanns}.
Fig.~\ref{fig:js_computing2} breaks down the computation latency, with over 40\% from distance calculations, 5\% from memory access, and the remaining 50\% from the data sorting and management of HNSW algorithm.
The results highlight that the extensive computational operations of ANNS implemented in interpreted languages have become a significant performance bottleneck.
Therefore, computational cost is one of the key limitations of in-browser ANNS.

\parabf{Opportunities and Challenges}. 
A potential solution is adopting Wasm~\cite{wasm}, a binary instruction format that provides near-native computing speeds in browsers, significantly improving performance over JavaScript for compute-intensive tasks.
However, employing Wasm in in-browser ANNS presents challenges due to its strict memory constraints and sandboxed isolation.
First, Wasm is limited by 32-bit addressing~\cite{wasm_memory_limit}, allowing a maximum of 4GB of memory, making it challenging for in-browser ANNS to scale with the size of personalized datasets.
Second, due to the strict sandboxing, Wasm lacks direct access to external storage, like the IndexedDB API, further complicating its use for in-browser ANNS.

\subsubsection{High Overhead from External Storage Access}

We conduct experiments under limited memory conditions using a subset of the Wiki dataset containing 50,000 items (Wiki-50k). 
Wiki-50k is the largest dataset that Mememo can handle without crashing under reduced memory. 
We gradually decrease the memory-data ratio and record the end-to-end query latency. 
For instance, with a memory-data ratio of 90\%, the memory can store up to 90\% of the data.

\begin{figure}[t!]
\centering
\begin{subfigure}[t]{0.22\textwidth}
    \centering
    \includegraphics[width=\textwidth]{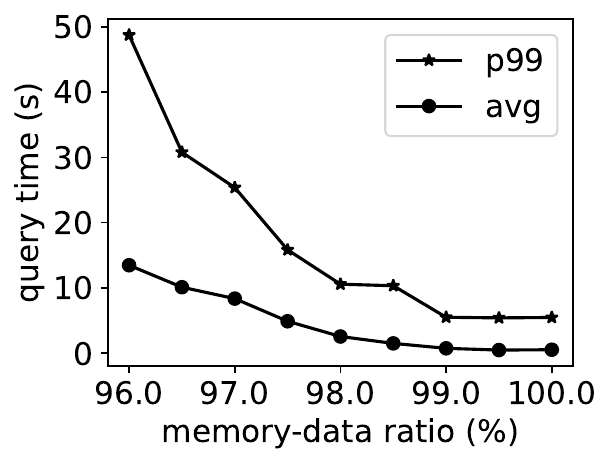}
    \caption{End-to-end query latency with limited memory.}
    \label{fig:disk_access1}
\end{subfigure}
\hfill
\begin{subfigure}[t]{0.23\textwidth}
    \centering
    \includegraphics[width=\textwidth]{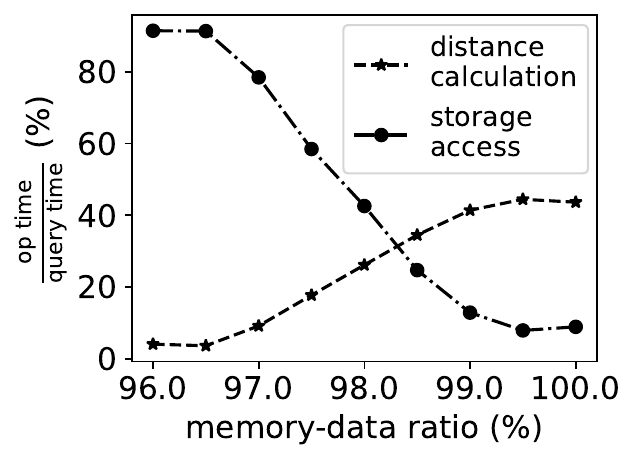}
    \caption{Bottleneck of queries with limited memory.}
    \label{fig:disk_access2}
\end{subfigure}

\vspace{-1em}
\caption{Overhead from external storage access.}
\vspace{-1.5em}

\end{figure}

Fig.~\ref{fig:disk_access1} shows that query latency rises rapidly as the memory-data ratio decreases slightly. 
When this ratio drops below 98\%, the P99 query latency exceeds 10 seconds, leading to unacceptable response latency for browsers.
At ratios under 96\%, the P99 latency surpasses 40 seconds, indicating that Mememo is unusable on memory-constrained browsers and devices.

Fig.~\ref{fig:disk_access2} also shows the breakdown of operations during the query process. 
When the memory-data ratio falls below 98\%, the time spent on external storage access exceeds that of distance calculations, and accounts for over 80\% at the ratio of 96\%, becoming the primary performance bottleneck.

\begin{figure}[t!]
\centering
\begin{subfigure}[t]{0.23\textwidth}
    \centering
    \includegraphics[width=\textwidth]{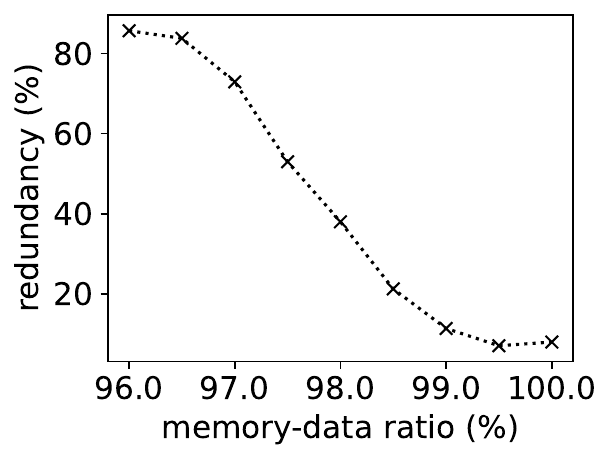}
    \caption{Redundancy rate of the prefetch strategy.}
    \label{fig:disk_opportunity1}
\end{subfigure}
\hfill
\begin{subfigure}[t]{0.22\textwidth}
    \centering
    \includegraphics[width=\textwidth]{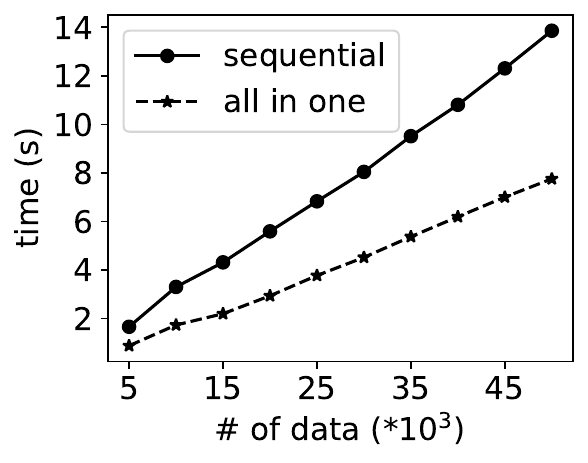}
    \caption{Latency of different loading strategies in IndexedDB.}
    \label{fig:disk_opportunity2}
\end{subfigure}
\vspace{-1em}
\caption{Optimization opportunity of storage access.}
\vspace{-1.5em}
\end{figure}

We further investigate the causes of high external storage access overhead.
Fig.~\ref{fig:disk_opportunity1} shows the redundancy rate ($R$) of the prefetch strategy, which is calculated by Equation~\ref{eq:redundancy},
\begin{equation}
    R = 1 - \frac{\#_{\text{hit}}}{(\#_{\text{disk access}} \times \#_{\text{prefetch size}})}
\label{eq:redundancy}
\end{equation}
where $\#_{\text{hit}}$ is the count of data hit in memory, $\#_{\text{disk access}}$ is the number of storage accesses, and $\#_{\text{prefetch size}}$ is the number of fetched data per access.

Fig.~\ref{fig:disk_opportunity1} indicates that when the memory-data ratio drops below 98\%, the redundancy rate exceeds 50\%, which means over half the loaded data is unused and causes unnecessary delay. 
This issue demonstrates the failure of the heuristic prefetch strategies of Mememo when the memory-data ratio decreases.
Therefore, we need more effective methods to reduce external disk accesses and minimize redundant loading across different memory-data ratios.

\parabf{Opportunities and Challenges}. 
Continuous IndexedDB accesses can cause high latency~\cite{indexeddb_slow}. 
An optimization opportunity is to maximize the effective data retrieved per disk access to reduce access frequency.

Fig.~\ref{fig:disk_opportunity2} compares IndexedDB latency for different loading strategies.
Sequential loading involves \textit{n} separate disk accesses for single items, while all-in-one loading retrieves \textit{n} items at one access. 
Results show that all-in-one loading is about 45.4\% faster than sequential loading, highlighting the overhead of transaction creation with each IndexedDB access.

However, the challenge lies in reducing IndexedDB access frequency effectively across different memory-data ratios. 
There is a trade-off between the amount of data loaded per access and redundancy. 
The queried vectors are unpredictable in the HNSW algorithm because the distance to a vector must be calculated before determining whether its neighbors should be queried.
Therefore, larger data loads per access can lead to less effective prefetching.

\subsubsection{Inefficient Use of Memory Resources}

Existing in-browser ANNS engines~\cite{wang2024mememo,voy,semanticfinder} overlook the optimization of memory utilization and rely mainly on a predefined memory threshold, which cannot adapt to different browsers and devices automatically.
The static threshold may lead to excessive memory usage during queries in memory-constrained devices, thereby affecting other browser functionalities and degrading user experience of web apps.

\parabf{Opportunities and Challenges.} 
As shown in Fig.~\ref{fig:disk_access1}, slightly reducing the memory-data ratio may not significantly affect query latency, suggesting the existence of an optimal memory threshold (the ratio of 99\% in Fig.~\ref{fig:disk_access1}). 
The opportunity lies in adaptively calculating this optimal memory threshold during the initialization stage of in-browser ANNS engine to reduce memory footprint while maintaining the query latency.
However, this threshold is influenced by disk access strategies, browsers, and devices, making it challenging to efficiently determine the optimal memory threshold.

\section{\sysname}

This section presents the overview and detailed design of \sysname.
We explain how the three-tier data management overcomes Wasm's address space limits, environment isolation, and execution model differences. 
We also discuss how \sysname minimizes IndexedDB accesses through lazy external storage access and how to determine the optimal memory threshold while ensuring the query latency.

\subsection{Overview}

\begin{figure}[t!]
\begin{center}
\includegraphics[width=0.48\textwidth]{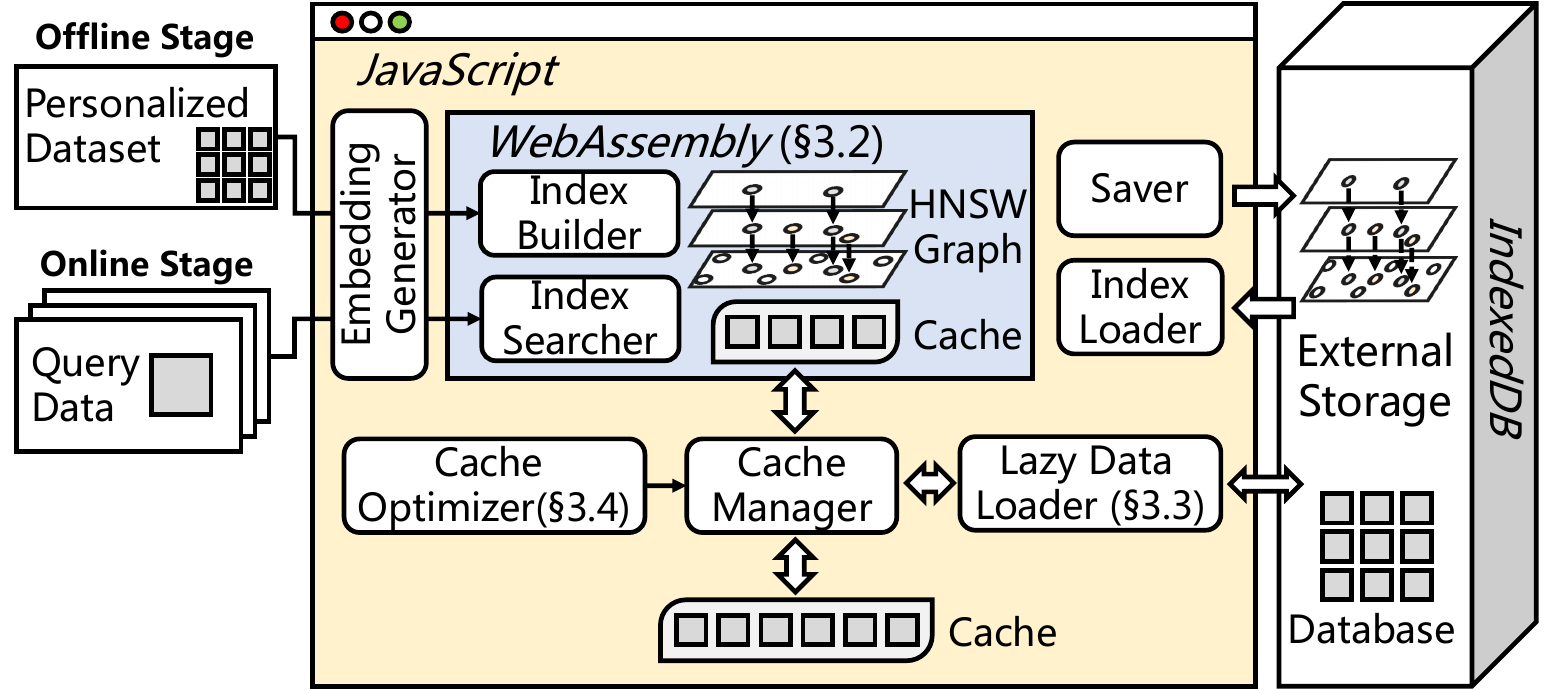}
\caption{Overview of \sysname.}
\label{fig:webrag}
\end{center}
\vspace{-1.8em}
\end{figure}

To comprehensively address the limitations of in-browser ANNS, we propose \sysname, which is designed to optimize query latency and memory utilization.
Fig.~\ref{fig:webrag} shows an overview of \sysname.

\parabf{Offline Indexing Construction}. The \sysname system starts by taking the user's personalized dataset as input. 
Using personalized text data and semantic embeddings, \sysname constructs an efficient indexing graph of HNSW.
The indexing graph, along with the text data and embeddings, is stored in IndexedDB for long-term storage.
Browsers provide mechanisms like service worker~\cite{service_worker} that enable the construction process to be conducted \emph{offline}.
Once the indexing graph is constructed, it can be loaded and reused for future queries.

\parabf{Online Data Query}. Online query latency is a primary focus for ANNS engines~\cite{jin2024ragcache,tian2024fusionanns,wang2024mememo} and directly impacts user experience.
The online query process of \sysname is divided into initialization and subsequent query stages.

During the initialization stage, \sysname loads the HNSW indexing graph into Wasm memory via the index loader and performs cache optimization to determine the optimal memory threshold, ensuring minimal memory usage without compromising query performance.
At the query stage, \sysname waits for query requests. Once a query is received, \sysname generates a query vector and then the index searcher performs a search on the HNSW graph based on the query vector.

\subsection{Three-Tier Data Management}
\label{subsec:opt1}

\sysname uses Wasm to perform HNSW construction and queries for enhanced computational performance. 
To address Wasm's limited memory and lack of external storage access, \sysname employs a three-tier data management mechanism with Wasm, JavaScript, and IndexedDB to manage data caching, data exchange, and coordination between synchronous and asynchronous execution models.

\parabf{Data Caching.}
When a browser's available memory exceeds Wasm's capacity, Wasm's limited memory would be a bottleneck for data cache size during the query process, causing more frequent accesses to external storage and degrading performance. 
To mitigate this, \sysname introduces a JavaScript cache between the Wasm cache and IndexedDB storage, forming a three-tier hierarchy. 
The first-tier cache is within Wasm, storing frequently accessed vectors in its limited memory. 
The second-tier cache is in the JavaScript environment, which holds more data when the Wasm cache is insufficient, thereby reducing the external storage access frequency.
The third tier is the external storage, where all data is kept within IndexedDB for retrieval as needed.

\parabf{Data Exchange.}
Wasm and IndexedDB cannot directly access each other due to their isolated environments. 
To facilitate data exchange, \sysname introduces JavaScript APIs as intermediaries. 
Specifically, JavaScript provides \textit{get()} and \textit{store()} APIs for Wasm to load missing vectors and write vectors to IndexedDB.
When Wasm encounters a data miss, it notifies JavaScript to check the second-tier cache via \textit{get()} API.
If the data is absent in the second-tier cache, it will be fetched from IndexedDB by JavaScript.
Once the data is loaded into memory, JavaScript will return it to Wasm, completing the \textit{get()} API calls.
If the first-tier cache needs to evict data, Wasm can move it to the second-tier cache or external storage via the \textit{store()} API provided by JavaScript.

\begin{figure}[t!]
\begin{center}
\includegraphics[width=0.45\textwidth]{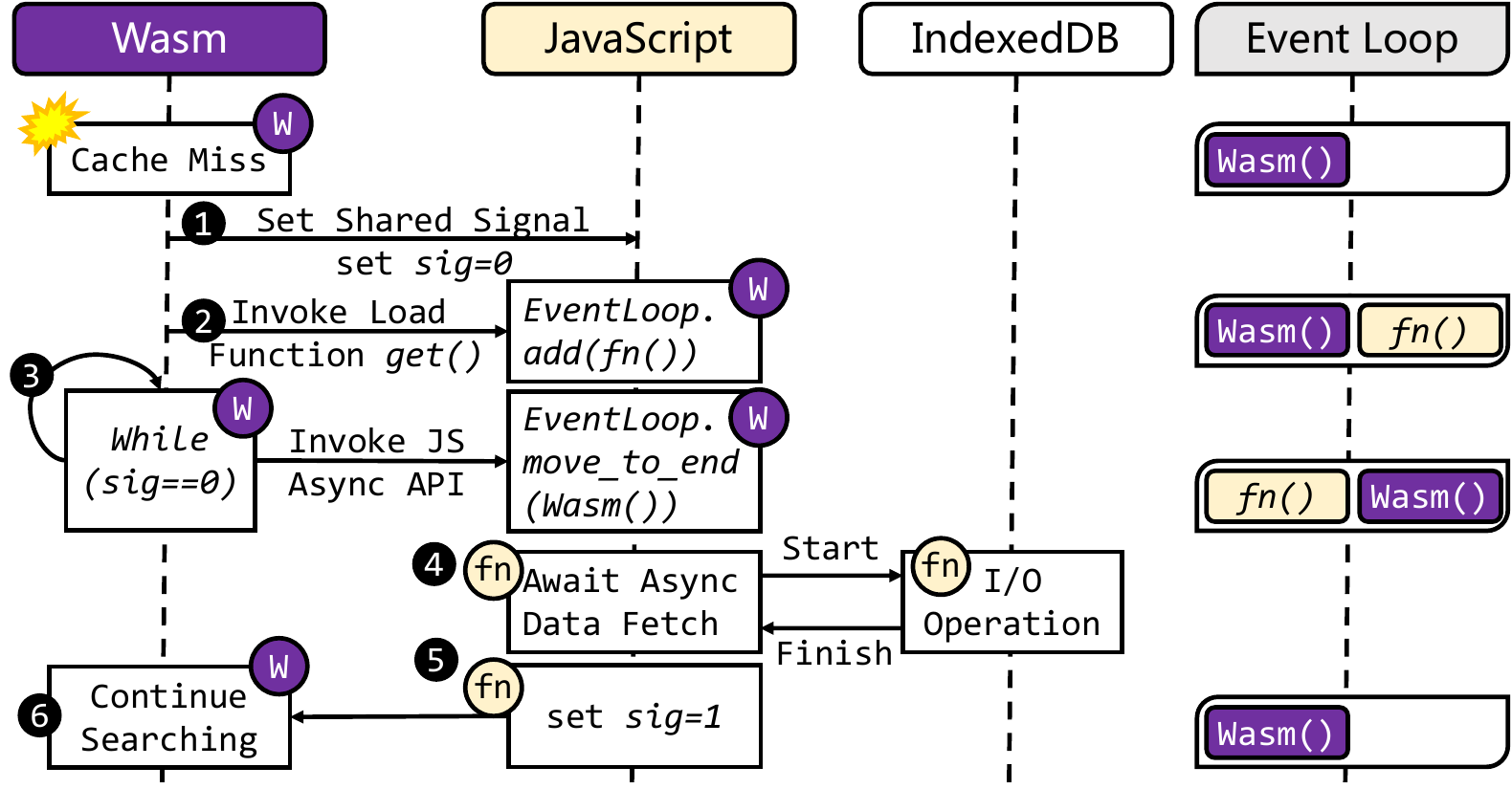}
\caption{Execution model coordination.}
\label{fig:execution_coordination}
\end{center}
\vspace{-1.8em}
\end{figure}

\parabf{Execution Model Coordination.}
Wasm and IndexedDB have fundamentally different execution models.
Wasm supports only synchronous execution, while IndexedDB operates asynchronously. 
This means that when a cache miss occurs in Wasm, it cannot asynchronously wait for the I/O operations of IndexedDB to complete. 
To address the difference, \sysname uses JavaScript as a bridge to connect Wasm's synchronous model with IndexedDB's asynchronous model.
The detailed coordination process is shown in Fig.~\ref{fig:execution_coordination}.

\blackcircle{1} A shared signal $sig$ is set, indicating the loading operation is pending.
\blackcircle{2} The \textit{get()} API is invoked to retrieve data. A new task of asynchronously loading data (\textit{fn()}) from IndexedDB via JavaScript is pushed into the end of the event loop of JavaScript.
\blackcircle{3} If the Wasm task is at the top of the event loop, Wasm continually monitors the shared signal $sig$.
If signal $sig$ has not changed, Wasm calls the asynchronous wait API provided by JavaScript to append the Wasm task to the end of the event loop, thereby freeing up computation during the wait.
\blackcircle{4} The data loading task of JavaScript moves to the top of event loop and can be executed to handle asynchronous I/O operations with IndexedDB.
\blackcircle{5} Once the data is fetched from IndexedDB, JavaScript sets the shared signal $sig$ as completion and passes it to the Wasm memory.
Then the data loading task is finished.
\blackcircle{6} The Wasm task is executed and detects the change of $sig$, and then resumes its computation.

Through the three-tier data management hierarchy, \sysname utilizes JavaScript as an intermediary cache, a data exchange bridge, and a connector between Wasm's and IndexedDB's execution models. 
This approach effectively overcomes Wasm's memory limitations and facilitates efficient data exchange between isolated environments with different execution models, thereby achieving Wasm acceleration for the query process.

\subsection{Lazy External Storage Access}
\label{subsec:opt2}

\begin{algorithm}[t!]
\caption{SEARCH-LAYER-WITH-PHASED-LAZY-LOADING}
\begin{algorithmic}[1]
\Statex \textbf{Input:} query element $q$, entry points $ep$, number of nearest elements to return $ef$, layer number $lc$.
\Statex \textbf{Output:} $ef$ closest neighbors to $q$.
\State $v \gets ep$ \textbf{// set of visited elements}
\State $C \gets ep$ \textbf{// set of candidates}
\State $W \gets ep$ \textbf{// dynamic list of found nearest neighbors}
\State $L \gets \emptyset$ \textbf{// candidates for lazy loading} \Comment{lazy}
\While{true}  \Comment{lazy}
    \color{black}
    \While{$|C| > 0$}
        \State $c \gets$ extract nearest element from $C$ to $q$
        \State $f \gets$ get furthest element from $W$ to $q$
        \If{distance($c$, $q$) $>$ distance($f$, $q$)}
            \State \textbf{break} \textbf{// all elements in $W$ are evaluated}
        \EndIf
        \For{each $e \in$ neighbourhood($c$) at layer $lc$}
            \If{$e \notin v$}
                \State $v \gets v \cup e$
                
                \If{ $e$ \textbf{not in} memory} \Comment{lazy}
                    \State $L \gets L \cup e$ \Comment{lazy}
                    \State \textbf{continue} \Comment{lazy}
                \EndIf
                
                \color{black}
                \If{ $e$ should be a candidate}
                    \State $C \gets C \cup e$
                    \State $W \gets W \cup e$
                    \If{$|W| > ef$}
                        \State remove furthest element in $W$
                    \EndIf
                \EndIf
            \EndIf
        \EndFor
        
        \If{$|L| > ef$} \Comment{intra-layer}
            \State \textbf{break} \Comment{lazy}
        \EndIf
    \color{black}
    \EndWhile

    \If{$|L| > 0$} \Comment{inter-layer}
        \State load elements in $L$ to memory \Comment{lazy}
        \For{each $e \in L$} \Comment{lazy}
            \If{ $e$ should be a candidate} \Comment{lazy}
                \State $C \gets C \cup e$ \Comment{lazy}
                \State $W \gets W \cup e$ \Comment{lazy}
                \If{$|W| > ef$} \Comment{lazy}
                    \State remove furthest element in $W$ \Comment{lazy}
                \EndIf
            \EndIf
        \EndFor
    \Else \Comment{lazy}
        \State \textbf{break} \Comment{lazy}
    \EndIf
\EndWhile
\color{black}
\State \Return $W$
\end{algorithmic}
\label{alg:searchlayer}
\end{algorithm}

Heuristic prefetching algorithms can lead to considerable redundant data loading, reducing the efficiency of external storage access. 
Given the high overhead of accessing external storage, it is crucial to minimize redundancy while maximizing the amount of data loaded per access.

\parabf{Completely Lazy Loading.} An extreme approach is to ignore all the data that should be queried but is not in memory during the query process.
Once the in-memory query is completed, all ignored data is loaded into memory with a single IndexedDB access to continue querying.
This completely lazy loading approach ensures that all data fetched from IndexedDB is required to be queried, thus eliminating redundant external data access.
However, this approach is unsuitable for the query process of HNSW for two reasons.
First, HNSW requires identifying the correct entry point for each layer.
If an entry point is ignored due to a cache miss, queries in subsequent layers will start from the wrong entry points, resulting in inaccurate results.
Second, calculating the distance to a queried vector is necessary for HNSW to decide whether its neighbors should be queried. 
Ignoring a queried vector might cause all its neighbors to be overlooked, thereby increasing the risk of following an incorrect query path.

\parabf{Phased Lazy Loading.} 
Building on the complete lazy loading approach, \sysname applies timing restrictions on lazy data loading to prevent excessive redundant computations along incorrect query paths.
The timing of data loading is determined by two key observations.
First, within the search space of a layer, the entry points of the next layer will be accurate if all necessary vectors are loaded and queried by the end of the search in that layer.
Second, when the list of ignored vectors exceeds the \textit{ef} parameter, which specifies the size of the candidate list for queries, the ignored list definitely includes extra vectors that do not require querying and neighbor evaluation.

Based on these observations, \sysname employs phased lazy loading, as shown in Algorithm~\ref{alg:searchlayer}.
For the \emph{inter-layer} query phase, after completing the search at each layer, any remaining lazily loaded vectors will be loaded in a single disk access, allowing the search to continue.
This process repeats until the search is completed with no remaining ignored vectors, ensuring correct entry points of the next layer.
For the \emph{intra-layer} query phase, if the list of ignored vectors exceeds the \textit{ef} parameter, all the vectors in the lazily loaded list will be loaded in a single disk access transaction. 
This process prevents the unlimited growth of the lazily loaded list, avoiding the loading of excessive redundant vectors.

The phased lazy load mechanism in \sysname minimizes external storage access by lazily loading data only when necessary. 
It ensures correct search paths by maintaining accurate entry points for each layer and controlling the size of the lazily loaded vector list, thereby eliminating excessive redundant loading.

\subsection{Heuristic Cache Size Optimization}
\label{subsec:opt3}

Due to the limited memory available to web apps, excessive memory usage by ANNS may disrupt other browser functionalities like user interactions and negatively affect the user experience. 
Therefore, \sysname aims to minimize memory usage of ANNS while maintaining query latency.

However, it is challenging to adaptively determine the optimal memory threshold, which is closely associated with the prefetch strategy, disk access strategy, browser, and device.

Our insight is to treat the query process as a black box. 
Starting with the maximum memory, we gradually reduce the memory limitation and observe changes in query latency to identify the optimal memory threshold.
Therefore, the problem lies in determining the appropriate step size for memory reduction.

\parabf{Na\"{\i}ve Method.}
A na\"{\i}ve method is to reduce memory with a fixed step size, but choosing the right size is challenging.
A small step size may require excessive query tests to find the optimal threshold, while a large step size might skip effective memory sizes and significantly increase the latency of query tests when the memory size drops below the optimal threshold.

\parabf{Heuristic Method}
To heuristically determine the step size for memory reduction, we model query latency under various memory sizes, as shown in Equation~\ref{eq:query_time},

\vspace{-1em}
\begin{align}
    T_{\text{query}} & = T_{\text{in-mem}} + T_{\text{db} } \notag \\
    & = |Q| \cdot t_{\text{in-mem}} + n_{db} \cdot t_{db}
\label{eq:query_time}
\end{align}

\revise{where $|Q|$ is the number of data items visited during one HNSW search, $Q$ denotes the ordered sequence of visited items (i.e., search path),}
$ t_{\text{in-mem}} $ is the time required for in-memory computations for each queried vector, $ n_{db} $ is the number of disk accesses, and $ t_{db} $ is the time taken for a single IndexedDB access.

Changes in memory size mainly affect $n_{db}$, ultimately impacting $T_{\text{query}}$. 
Therefore, we can focus on the relationship between the number of items that can be stored in memory $ n_{\text{mem}} $ and $ n_{db} $.
This relationship is mainly affected by the data fetch strategy from IndexedDB to memory.

\paraf{With random data fetching}, it can be proven that $n_{db}$ decreases linearly with an increase in $n_{\text{mem}}$, satisfying Equation~\ref{eq:random_cache},

\begin{equation}
n_{db} = 
\begin{cases}
\frac{1-|Q|}{N-1} \cdot n_{\text{mem}} + \frac{N\cdot|Q|-1}{N-1}, & \text{if } n_{\text{mem}} < N \\
1, & \text{if } n_{\text{mem}} \geq N
\end{cases}
\label{eq:random_cache}
\end{equation}
where $N$ is the number of items in the dataset.

\begin{myproof}
Consider a query that needs to access $|Q|$ vectors from a dataset containing $N$ items, with the memory capable of holding $n_{\text{mem}}$ vectors.

\textbf{(1)} When $n_{\text{mem}} \geq N$, all vectors can be loaded into memory in a single IndexedDB access.

\textbf{(2)} When $n_{\text{mem}} = M < N$ and the memory is initially empty:

For the first queried vector $D_1$, a cache miss occurs, requiring one IndexedDB access to load $D_1$ and $M-1$ other random vectors into memory. Thus, the memory contains $D_1$ and $M-1$ other random data items.

For the $i$-th queried vector (denoted as $D_{i}$, $i \geq 2$), which is different from $D_{i-1}$, the probability that $D_{i}$ is in memory is given by
$P_{\text{hit}} = \frac{M - 1}{N - 1}.$
Since the memory contains $D_{i-1}$ and $M-1$ data items randomly selected from the dataset. If a cache miss occurs, another IndexedDB access is performed to load $D_{i}$ and $M-1$ other random data items.

Thus, the first query always results in an IndexedDB access. For the $i$-th query ($2 \leq i \leq |Q|$), the probability of a cache miss per query is, 
$1 - P_{\text{hit}} = \frac{N - M}{N - 1}.$
Therefore, the total number of IndexedDB accesses is 
$n_{db} = 1 + (|Q| - 1) \cdot \frac{N - M}{N - 1},$
which can be simplified to
$n_{db} = \frac{1-|Q|}{N-1} \cdot n_{\text{mem}} + \frac{N \cdot |Q| - 1}{N-1}.$

\end{myproof}

\paraf{With optimal data fetching}, each IndexedDB access can prefetch the next $n_{\text{mem}}$ vectors in the query path into memory, demonstrating that $n_{db}$ is inversely proportional to $n_{\text{mem}}$, as shown in Equation~\ref{eq:best_cache}.

\begin{equation}
n_{db} =
\begin{cases} 
\left\lceil \frac{|Q|}{n_{\text{mem}}} \right\rceil, & \text{if } n_{\text{mem}} < |Q| \\
1, & \text{if } n_{\text{mem}} \geq |Q|
\end{cases}
\label{eq:best_cache}
\end{equation}

\begin{myproof}
Consider a query that needs to access $|Q|$ vectors from a dataset containing $N$ items, with the memory capable of holding $n_{\text{mem}}$ vectors.
With optimal data fetching, each disk access can always prefetch the next $n_{\text{mem}}$ vectors in the query path Q.

\textbf{(1)} When $n_{\text{mem}} \geq |Q|$, the memory can hold all the vectors in the query path within a single IndexedDB access.

\textbf{(2)} When $n_{\text{mem}} < |Q|$, the minimum number of IndexedDB accesses needed is determined by how many full sets of $n_{\text{mem}}$ vectors can be loaded to cover all $|Q|$ vectors. This is given by 
$
n_{db} = \left\lceil \frac{|Q|}{n_{\text{mem}}} \right\rceil.
$
\end{myproof}

\begin{algorithm}[t!]
\caption{APPROXIMATING-CURVE-OF-REAL-FETCHING-STRATEGY}
\begin{algorithmic}[1]
    \Function{OPTIMIZE\_MEMORY\_SIZE}{$C_{0}$, $p$, $T_{\theta}$}
        \State \textbf{// $C_{0}$ is maximum memory size}
        \State \textbf{// $p$ and $T_{\theta}$ are the percentage and absolute thresholds}
        \State $C_{best} \gets C_{0}$ \textbf{// best memory size}
        \State $C_{test} \gets C_{0}$ \textbf{// memory size to be evaluated}
        \While{$0 < C_{test} \leq C_{0}$}
            \State $n_{db}$, $n_{Q}$, $T_{query}$, $t_{db}$ $\gets \text{QUERY\_TEST()}$
            \State \textbf{// $n_{db}$ is number of disk access, $n_{Q}$ is length of query path, $T_{query}$ is total query time, $t_{db}$ is time for single disk access}
            \State $\theta \gets \text{GET\_THETA($p$, $T_{\theta}$, $T_{query}$, $t_{db}$)}$
            \If{$n_{db} > \theta$}
                \State \textbf{Break // Over the threshold}
            \Else
                \State $C_{best} \gets C_{test}$
            \EndIf
            \State $k \gets \frac{n_{Q} - n_{db}}{1-C_{test}}$
            \State $C_{test} \gets \left\lceil \frac{\theta - n_{Q}}{k} + 1 \right\rceil$
        \EndWhile
        \State \Return $C_{best}$
    \EndFunction
    \\
    \Function{GET\_THETA}{$p$, $T_{\theta}$, $T_{query}$, $t_{db}$}
        \State \textbf{// $p$ and $T_{\theta}$ are the percentage and absolute thresholds}
        \State $\theta \gets max\left(\frac{p \cdot T_{query}}{t_{db}}, \frac{T_{\theta}}{t_{db}}\right)$
        \State \Return $\theta$.
    \EndFunction
\end{algorithmic}
\label{alg:cache_opt}

\end{algorithm}

\begin{figure}[t!]
\begin{center}
\includegraphics[width=0.45\textwidth]{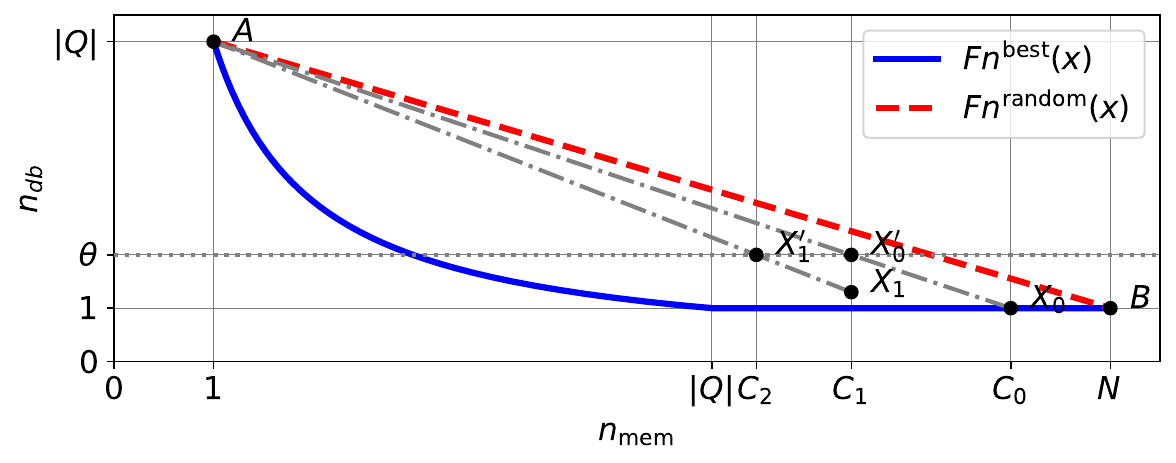}
\vspace{-1em}
\caption{Approximating the curve of real fetching strategy.}
\label{fig:cache_opt}
\end{center}
\vspace{-2em}
\end{figure}

\paraf{Approximating the curve of real fetching strategy.}
We hypothesize that the real fetching strategy results in fewer IndexedDB accesses than random fetching. Therefore, the curve should lie between $\text{Fn}^{\text{optimal}}(x)$ and $\text{Fn}^{\text{random}}(x)$, as shown in Fig.~\ref{fig:cache_opt}. By setting a threshold $\theta$ for $n_{db}$, we can start with the maximum memory size $C_0$ and iteratively reduce memory to find the optimal memory size under the real fetching strategy.

Algorithm~\ref{alg:cache_opt} shows the overall memory optimization process.
\sysname evaluates the number of IndexedDB accesses at the maximum memory size $C_0$ through query testing, represented as point $X_0$. If the accesses at $C_0$ exceed the threshold $\theta$, \sysname retains the memory size at $C_0$.
Otherwise, the next memory size $C_1$ is found by intersecting the line from $X_0$ to endpoint $A$ with $y = \theta$. 
For each subsequent memory size $C_i$ ($i \geq 1$), if accesses exceed $\theta$, $C_{i-1}$ will be selected as the optimal memory size.

This algorithm can rapidly approximate the curve of the real fetching strategy, based on two observations.
First, the gradient near the optimal memory size is relatively gentle, allowing lines connecting to the extreme point to quickly converge on the optimal memory size.
Second, as the algorithm progresses, the memory reduction steps will decrease, while the absolute value of the gradient of the line connecting to the extreme point increases and the database accesses in query tests approach $\theta$.

\paraf{Setting of $\theta$.}
$\theta$ can be set using two methods based on Equation~\ref{eq:query_time}. 
The first method uses a percentage parameter $p$ to ensure external storage access time remains below a specific ratio of $T_{\text{query}}$, calculated as $\theta = p \times \frac{T_{\text{query}}}{t_{db}}$.
\revise{Higher $p$ reduces memory but increases latency, and vice versa.}
The second method sets an absolute time limit to ensure the total time of IndexedDB access does not exceed a certain threshold, calculated as $\theta = \frac{T_{\theta}}{t_{db}}$, where $T_{\theta}$ is the acceptable IndexedDB access time, such as 100ms. 
\sysname incorporates both methods for setting $\theta$.

\paraf{Rollback of memory size.} 
The memory optimization algorithm tracks a sequence of memory sizes $\{C_0, C_1, \ldots\}$ and their corresponding $\theta$ values. 
During queries, if the IndexedDB access count at $C_i$ exceeds the corresponding $\theta$, \sysname would roll back the memory size to $C_{i-1}$. 
The rollback process continues until $C_0$ is reached, ensuring a rapid return to a memory size that maintains query performance when fluctuations occur.

Through heuristic cache size optimization, \sysname can maintain the query latency while minimizing memory footprint.

\begin{table*}[t!]
\centering
    \caption{P99 query time (ms) across heterogeneous devices and browsers for 100 queries with unrestricted memory usage.}
    \vspace{-1em}
    \label{tab:e2e_query}
    \resizebox{\textwidth}{!}{%
    
    \begin{tabular}{lrrrrrrrrr}
        \toprule
        \multirow{2}{*}{\textbf{Device/Browser}} & \multicolumn{2}{c}{\textbf{Wiki-60k (901MB)}} & \multicolumn{1}{c}{\textbf{Wiki-480k (7.5GB)}} & \multicolumn{2}{c}{\textbf{Arxiv-1k (4MB)}} & \multicolumn{2}{c}{\textbf{Arxiv-120k (460MB)}} & \multicolumn{2}{c}{\textbf{Finance-13k (214MB)}}      \\
      \cmidrule(lr){2-3} \cmidrule(lr){4-4} \cmidrule(lr){5-6} \cmidrule(lr){7-8} \cmidrule(lr){9-10}
  & \textbf{Meme./\sysname} & \textbf{Boost} & \textbf{Meme./\sysname} & \textbf{Meme./\sysname} & \textbf{Boost} & \textbf{Meme./\sysname} & \textbf{Boost} & \textbf{Meme./\sysname} & \textbf{Boost} \\
        \midrule
        \textbf{Linux/Chrome} &  7636.30 / \textbf{13.49} & 566.28$\times$  &  N/A / \textbf{23.46}  &  7.99 / \textbf{1.54} & 5.17$\times$  &  52.77 / \textbf{16.12} & 3.27$\times$  &  40.31 / \textbf{8.58} & 4.70$\times$ \\
        \textbf{Linux/Firefox} &  3865.22 / \textbf{26.38} & 146.52$\times$  &  N/A / \textbf{45.92}  &  10.60 / \textbf{2.30} & 4.61$\times$  &  170.74 / \textbf{21.62} & 7.90$\times$  &  33.80 / \textbf{17.18} & 1.97$\times$ \\
        \midrule
        \textbf{Mac/Chrome} &  11975.20 / \textbf{16.10} & 743.80$\times$  &  N/A / \textbf{30.40}  &  5.70 / \textbf{1.50} & 3.80$\times$  &  44.80 / \textbf{13.60} & 3.29$\times$  &  39.10 / \textbf{8.70} & 4.49$\times$ \\
        \textbf{Mac/Firefox} &  3023.00 / \textbf{43.00} & 70.30$\times$  &  N/A / \textbf{72.00}  &  9.00 / \textbf{4.00} & 2.25$\times$  &  69.00 / \textbf{30.00} & 2.30$\times$  &  33.00 / \textbf{22.00} & 1.50$\times$ \\
        \textbf{Mac/Safari} &  3107.00 / \textbf{16.00} & 194.19$\times$  &  N/A / \textbf{28.00}  &  4.00 / \textbf{2.00} & 2.00$\times$  &  61.00 / \textbf{14.00} & 4.36$\times$  &  20.00 / \textbf{8.00} & 2.50$\times$ \\
        \midrule
        \textbf{Win/Chrome} &  16937.10 / \textbf{23.40} & 723.81$\times$  &  N/A / \textbf{39.10}  &  10.50 / \textbf{2.60} & 4.04$\times$  &  80.70 / \textbf{22.00} & 3.67$\times$  &  66.90 / \textbf{11.60} & 5.77$\times$ \\
        \textbf{Win/Firefox} &  11975.00 / \textbf{42.00} & 285.12$\times$  &  N/A / \textbf{65.00}  &  16.00 / \textbf{3.00} & 5.33$\times$  &  132.00 / \textbf{32.00} & 4.12$\times$  &  60.00 / \textbf{21.00} & 2.86$\times$ \\
        \textbf{Win/Firefox} &  11975.00 / \textbf{42.00} & 285.12$\times$  &  N/A / \textbf{65.00}  &  16.00 / \textbf{3.00} & 5.33$\times$  &  132.00 / \textbf{32.00} & 4.12$\times$  &  60.00 / \textbf{21.00} & 2.86$\times$ \\
        \bottomrule
    \end{tabular}
}
\vspace{-1em}
\end{table*}

\begin{table}[t!]
\centering
    \caption{Ablation experiments on P99 query time (ms) with limited memory.}
    \label{tab:ablation}

    \vspace{-1em}
    
    \resizebox{0.75\linewidth}{!}{%
    \begin{tabular}{c c c c c c}
        \cellcolor{gray!50} & \textbf{Mememo} & 
        \cellcolor{gray!100} & \textbf{\sysname-Base} & 
        \cellcolor{black} & \textbf{\sysname} \\
    \end{tabular}
    }
    
    \resizebox{0.9\linewidth}{!}{%

    \begin{tabular}{llrrrrr}
        \toprule
        \multirow{2}{*}{\textbf{\makecell[tl]{Device/\\Browser}}} & & \multicolumn{5}{c}{\textbf{Memory/Data Ratio (\%)}} \\
        \cmidrule(lr){3-7}
        & & \textbf{20\%} & \textbf{90\%} & \textbf{96\%} & \textbf{98\%} & \textbf{100\%} \\
        \midrule
        \multirow{3}{*}{\textbf{\makecell[tl]{Linux/\\Chrome}}}        & \cellcolor{gray!50} & N/A & N/A & 48743.38 & 10574.26 & 5472.33 \\
        & \cellcolor{gray!100} & 34216.27 & 1760.31 & 825.44 & 477.67 & \textbf{14.07} \\
        & \cellcolor{black} & \textbf{425.39} & \textbf{106.97} & \textbf{50.01} & \textbf{34.58} & 14.19 \\
        \cmidrule(lr){3-7}
        \multirow{3}{*}{\textbf{\makecell[tl]{Linux/\\Firefox}}}        & \cellcolor{gray!50} & N/A & N/A & 16917.14 & 5478.30 & 2500.22 \\
        & \cellcolor{gray!100} & 32705.08 & 1940.90 & 866.40 & 390.90 & \textbf{28.42} \\
        & \cellcolor{black} & \textbf{591.36} & \textbf{167.58} & \textbf{127.40} & \textbf{71.64} & 28.50 \\
        \midrule
        \multirow{3}{*}{\textbf{\makecell[tl]{Mac/\\Chrome}}}        & \cellcolor{gray!50} & N/A & N/A & N/A & 12563.10 & 7480.00 \\
        & \cellcolor{gray!100} & 42105.90 & 2369.40 & 1375.20 & 444.80 & 15.60 \\
        & \cellcolor{black} & \textbf{501.80} & \textbf{91.00} & \textbf{54.90} & \textbf{49.20} & \textbf{15.20} \\
        \cmidrule(lr){3-7}
        \multirow{3}{*}{\textbf{\makecell[tl]{Mac/\\Firefox}}}        & \cellcolor{gray!50} & N/A & N/A & 12267.00 & 4730.00 & 2475.00 \\
        & \cellcolor{gray!100} & 50334.00 & 3603.00 & 1417.00 & 662.00 & \textbf{43.00} \\
        & \cellcolor{black} & \textbf{1060.00} & \textbf{161.00} & \textbf{92.00} & \textbf{77.00} & 45.00 \\
        \cmidrule(lr){3-7}
        \multirow{3}{*}{\textbf{\makecell[tl]{Mac/\\Safari}}}        & \cellcolor{gray!50} & N/A & N/A & 26540.00 & 8041.00 & 4742.00 \\
        & \cellcolor{gray!100} & 39078.00 & 2185.00 & 915.00 & 480.00 & 16.00 \\
        & \cellcolor{black} & \textbf{688.00} & \textbf{98.00} & \textbf{61.00} & \textbf{44.00} & \textbf{16.00} \\
        \midrule
        \multirow{3}{*}{\textbf{\makecell[tl]{Win/\\Chrome}}}        & \cellcolor{gray!50} & N/A & N/A & 65895.50 & 23161.10 & 12212.80 \\
        & \cellcolor{gray!100} & 42374.80 & 3286.00 & 923.40 & 514.10 & 22.50 \\
        & \cellcolor{black} & \textbf{828.90} & \textbf{110.50} & \textbf{70.10} & \textbf{57.20} & \textbf{21.60} \\
        \cmidrule(lr){3-7}
        \multirow{3}{*}{\textbf{\makecell[tl]{Win/\\Firefox}}}        & \cellcolor{gray!50} & N/A & N/A & 55831.00 & 16301.00 & 8998.00 \\
        & \cellcolor{gray!100} & 124241.00 & 6453.00 & 3716.00 & 1708.00 & \textbf{39.00} \\
        & \cellcolor{black} & \textbf{2127.00} & \textbf{314.00} & \textbf{157.00} & \textbf{142.00} & 42.00 \\
        \bottomrule
    \end{tabular}

    }
\vspace{-1.5em}
\end{table}

\section{Experiments}

This section presents the implementation and experiments of \sysname.

\subsection{Implementation}

Based on the JavaScript implementation of HNSW from Mememo~\cite{wang2024mememo}, we implement HNSW in C++ and compile it into Wasm. 
The prototype of \sysname is developed using over 3,300 lines of C++ and TypeScript code. 
We also develop a web app with a user interface to validate the effectiveness of \sysname in browsers. 
Apart from the three key optimizations mentioned before, the prototype system includes the following enhancements.

\paraf{Text-embedding separation.} 
In in-browser ANNS engines, only embeddings are needed during queries, not the original text. 
Storing text-embedding pairs of data directly as key-value pairs can result in high memory overhead due to the length of the text. 
To address this, we assign a unique ID to each data item and manage texts and embeddings separately. This ID indexes both embeddings and texts, minimizing memory usage for each entry.

\paraf{Streaming data loading.}
In browser environments, loading user-uploaded personalized datasets or large HNSW index trees can exceed memory limits, risking browser crashes. 
To address this, the \sysname prototype system supports stream loading, enabling index trees and data files to be read and loaded in chunks.

\paraf{Cache eviction strategy.} 
For simplicity, the \sysname prototype uses FIFO as the eviction strategy for caches in both Wasm and JavaScript. 
Note that \sysname prototype provides a unified abstract interface for cache eviction, allowing the system to easily support various eviction algorithm implementations as pluggable modules.

\subsection{Setup}

\parabf{Dataset.}
We evaluate our approach using one general dataset (Wiki) and two domain-specific datasets (ArXiv and Finance).
\textbf{(1) Wiki}~\cite{wiki_dataset} is the dataset used in our measurements study. 
\revise{It contains 480,000 text passages and their 768-dimensional vector embeddings.}
The dataset is 7.5GB in size and widely used for evaluating the performance of ANNS~\cite{jin2024ragcache}.
\textbf{(2) Arxiv} is a collection of two datasets with sizes of 1k and 120k, containing the abstracts of arXiv machine learning papers. 
Following prior work~\cite{wang2024mememo,mememo_app}, these datasets are used to assess the in-browser ANNS engine's performance across varying scales.
\textbf{(3) Finance} is the FinDER dataset~\cite{linq2023finder}, open-sourced by LinQ, which includes more than 10k reports and financial disclosures. 
It contains a substantial corpus of domain-specific text, representing the scale of personalized and privacy-sensitive datasets.

\parabf{Baseline.}
We compare \sysname with the following two baselines.
For each baseline, we report the performance achieved without encountering out-of-memory errors.
\textbf{(1) Mememo}~\cite{wang2024mememo} is the SOTA in-browser ANNS engine, which is published in \emph{SIGIR'24}.
\textbf{(2) \sysname-Base} removes lazy loading and memory optimization but includes all other optimizations. 
It highlights the benefits of Wasm from Section~\ref{subsec:opt1}. 
The comparison with \sysname-Base underscores gains from techniques in Sections~\ref{subsec:opt2} and \ref{subsec:opt3}, eliminating any unfair comparisons caused by the differences in underlying implementations.

\parabf{Metrics.} 
We use the query latency of in-browser ANNS as a metric, following prior work~\cite{zhang2023fast,tian2024fusionanns}. 
Since neither Mememo nor our work optimizes the conversion of query text to embeddings, we measure end-to-end latency from inputting the query embeddings to receiving the search results. 
In each experiment setting, after an initial warm-up query, we record latency over 100 iterations and calculate the 99th percentile (P99) latency to represent the duration for most queries.

\parabf{Environment.}
We conduct experiments on a Linux desktop with an Intel Core i5-13400F processor (13th generation), a MacOS laptop with an Apple M2 processor, and a Windows laptop with an AMD R7-4800U processor. 
The browsers used for experiments involve Chrome (version 131.0), Firefox (version 133.0), and Safari (version 18, macOS only).

\subsection{Query Performance}

Table~\ref{tab:e2e_query} shows the query latency comparison when memory usage is unrestricted.
\sysname demonstrates a significant speedup across all evaluated conditions, with boosts ranging from $1.5\times$ to $743.8\times$.

For large datasets, \sysname shows more significant improvements. 
This is mainly because Mememo's heuristic prefetching strategy may fail to load all necessary data into memory, even without memory restrictions, leading to multiple external storage accesses. 
\sysname minimizes external storage access through a three-tier cache and lazy loading techniques, achieving a 70-744x performance boost on large datasets like Wiki, reducing retrieval times from seconds to around 10ms.
This enhancement transforms nearly unusable searches of Mememo into usable ones.

For smaller datasets, where external storage accesses are not involved, \sysname fully leverages the computing efficiency of Wasm, achieving a $2$--$5.33\times$ improvement on the smallest dataset, Arxiv-1k.

Additionally, \sysname expands the dataset size supported for retrieval in browsers by at least 8.5 times. 
Frequent external access and excessive memory usage cause Mememo to crash when handling datasets larger than Wiki-60k, whereas \sysname can complete the retrieval of the 7.5GB Wiki-480k dataset in 72ms.

\paraf{Impact of browser and device heterogeneity.}
\sysname consistently improves query speed across all browsers and devices.
Compared to Linux desktops with more powerful processors, \sysname achieves higher boost rates on Mac and Windows laptops, reaching a $743\times$ boost. 
Regarding the performance of \sysname across browsers, it achieves the lowest latency on Chrome, reaching as low as 1.5ms, while Firefox is the slowest, likely due to the different performance of their Wasm engines.

\subsection{Ablation Study}

We conduct ablation experiments on various devices with the Wiki-50k dataset, the largest dataset that Mememo can handle without crashing under reduced memory. 
We set memory equal to the dataset size and reduce the memory-data ratio incrementally. 
For example, a 90\% memory-data ratio means the memory holds up to 90\% of the data.
Results are shown in Table~\ref{tab:ablation}.

\parabf{Three-tier data management.}
Mememo becomes unusable with sub-minute level query times and even crashes with a memory-data ratio below 98\% due to exponential query time increases. 
\sysname-Base, with Wasm and three-tier caching, archives at least an order-of-magnitude improvement compared to Mememo.

\parabf{Lazy external storage access.}
The lazy loading mechanism of \sysname reduces external access at best efforts, further achieving at least an order-of-magnitude improvement compared to \sysname-Base at memory ratio below 90\%.
Moreover, \sysname can save much memory footprint within tolerable query performance, keeping sub-second query latency even at a 20\% memory ratio.
Besides, the lazy storage access mechanism does not introduce much overhead compared to \sysname-Base. When memory can hold all data (memory-data ratio=100\%), \sysname has a competitive performance to \sysname-Base.

\begin{table}[t!]
\centering
    \caption{Heuristic cache size optimization ($p=0.8, T_{\theta}=100ms$)}
    \label{tab:cache_opt}

    \vspace{-0.5em}
    
    \resizebox{\linewidth}{!}{%

    \begin{tabular}{lrrrr}
        \toprule
        \textbf{Device/} & \textbf{Init} & \textbf{Opt} & \textbf{Saved} & \textbf{P99 Query} \\
        \textbf{Browser} & \textbf{Mem. (MB)} & \textbf{Mem. (MB)} & \textbf{Mem. (MB)} & \textbf{Time (ms)} \\
        \midrule
        \textbf{Linux/Chrome} & 146.48 & 101.70 & 44.79 (31\%)& 170.22 \\
        \cmidrule(lr){1-5}
        \textbf{Linux/Firefox} & 146.48 & 104.23 & 42.25 (29\%)& 257.80 \\
        \midrule
        \textbf{Mac/Chrome} & 146.48 & 89.08 & 57.40 (39\%)& 186.30 \\
        \cmidrule(lr){1-5}
        \textbf{Mac/Firefox} & 146.48 & 111.65 & 34.83 (24\%)& 276.00 \\
        \cmidrule(lr){1-5}
        \textbf{Mac/Safari} & 146.48 & 113.49 & 33.00 (23\%)& 182.00 \\
        \midrule
        \textbf{Win/Chrome} & 146.48 & 124.68 & 21.81 (15\%)& 151.00 \\
        \cmidrule(lr){1-5}
        \textbf{Win/Firefox} & 146.48 & 136.59 & 9.89 (7\%)& 222.00 \\
        \bottomrule
    \end{tabular}

    }
\vspace{-1em}
\end{table}

\parabf{Heuristic cache size optimization.}
With $p = 0.8$ and $T_{\theta}=100\text{ms}$, as shown in Table~\ref{tab:cache_opt}, \sysname adaptively minimizes memory usage based on device and browser performance, saving 7\%-39\% memory while maintaining query times. 
Note that this optimization process only needs to run once at web app startup, because optimal memory usage depends mainly on the runtime environment.


\section{Related Work}

\parabf{Approximate nearest neighbor search algorithm.}
ANNS has become a critical component in modern information retrieval and AI systems, with diverse applications such as RAG and recommendation systems. 
In general, ANNS algorithms can be categorized into tree-based~\cite{annoy,li2023constructing}, hashing~\cite{johnson2019billion,pham2022falconn++}, quantization-based methods~\cite{ge2013optimized,jegou2010product}, and graph-based~\cite{hnsw}.
For in-browser ANNS~\cite{wang2024mememo}, graph-based algorithms are widely adopted for their SOTA performance regarding construction and query efficiency.
Following previous work, \sysname focuses on the graph-based HNSW algorithm and further optimizes the query latency at the system level.

\parabf{ANNS systems.}
With the increasing popularity of ANNS, recent work of ANNS systems aims to optimize performance across different environments, including cloud~\cite{zhang2023fast,tian2024fusionanns}, edge~\cite{seemakhupt2024edgerag}, and web browsers~\cite{wang2024mememo,voy,semanticfinder}.
\sysname specifically focuses on in-browser ANNS to enhance its integration with existing web apps and optimize query latency in browsers, making in-browser ANNS practical for more extensive datasets.

\section{Conclusion}

We present \sysname, an efficient in-browser ANNS engine that addresses computational, memory, and storage access challenges specific to web browsers. 
By leveraging Wasm, lazy loading, and heuristic memory optimization, WebANNS achieves up to two orders of magnitude faster P99 query latency, 40\% lower memory usage, and supports over $8 \times$ larger data size compared to the SOTA engine.
\sysname enables privacy-preserving, high-performance ANNS in web, making in-browser ANNS practical to incorporate into real-world web applications.

\section*{Acknowledgment}
This work was supported by the National Natural Science Foundation of China under the grant number 62325201.

\bibliographystyle{ACM-Reference-Format}
\balance
\bibliography{webrag}

\end{document}